\documentstyle[11pt,epsf,paspconf]{article}

\begin{document}

{\footnotesize \raggedleft FERMILAB-Conf-99/244-A \\}

\title{Cosmological Parameters and Power Spectrum from Peculiar Velocities}

\author{I. Zehavi}
\affil{NASA/Fermilab Astrophysics Group, Fermi National Accelerator 
Laboratory, Box 500, Batavia, IL 60510-0500, USA}

\author{A. Dekel}
\affil{Racah Institute of Physics, The Hebrew University,
Jerusalem 91904, Israel}

\begin{abstract}
The power spectrum of mass density fluctuations is evaluated 
from the Mark~III and the SFI catalogs of peculiar velocities by
a maximum likelihood analysis, using parametric models
for the power spectrum and for the errors. The applications to the 
two different data sets, using generalized CDM models with and without 
COBE normalization, give consistent results. The general result is a 
relatively high amplitude of the power spectrum, with
$\sigma_8 {\Omega_m}^{0.6} = 0.8 \pm 0.2$ at $90\%$ confidence.
Casting the results in the $\Omega_m - \Omega_\Lambda$ plane, yields
complementary constraints to those of the high-redshift supernovae,
together favoring a nearly flat, unbound and accelerating universe with 
comparable contributions from $\Omega_m$ and $\Omega_\Lambda$. Further 
implications on the cosmological parameters, arising from a joint analysis 
of the velocities together with small-scale CMB anisotropies and the 
high-redshift supernovae, are also briefly described.
\end{abstract}


\section{Introduction}
\label{sec:intro}
In the standard picture of cosmology, structure evolved from small
density fluctuations that grew by gravitational instability. These
initial fluctuations are assumed to have a Gaussian distribution
characterized by the power spectrum (PS). On large scales, the 
fluctuations are linear even at late times and still governed by the 
initial PS. The PS is thus a useful statistic for large-scale structure, 
providing constraints on cosmology and theories of structure formation. 
It is advantageous to estimate it from velocities, as these are
directly related to the {\it mass} density fluctuations, and are effected
by large scales and thus are approximately still linear.

In this work, we develop and apply a likelihood analysis for estimating 
the mass PS from peculiar velocities.
This method uses the ``raw'' peculiar velocities without additional 
processing, and thus utilizes much of the information content of the data. 
It also takes into account properly the measurement errors and the finite 
discrete sampling. The simplifying assumptions made are that the velocities 
follow a Gaussian distribution and that their correlations can be derived 
from the density PS using linear theory. 

We use the two comprehensive available peculiar velocity
data sets, the Mark~III catalog (Willick et al.\ 1997) and the SFI catalog
(Haynes et al.\ 1999a,b; Giovanelli, these proceedings). Mark~III samples
$\sim 3000$ galaxies within a distance of $\sim 70 h^{-1}{\rm Mpc}$ 
around us, and SFI consists of $\sim 1300$ spiral galaxies with a more uniform
spatial coverage in a similar volume. The typical relative distance errors 
of individual galaxies are $15-20\%$, and both catalogs are carefully 
corrected for systematic errors.
It is interesting to compare the results of the two catalogs, especially 
in view of apparent discrepancies in the appearance of the velocity fields 
(e.g., da Costa et al.\ 1996, 1998). 
We explore the cosmological implications of the velocities on their 
own and in conjunction with constraints derived from other types of data.

\section{Method}
\label{sec:method}
Given a data set ${\bf d}$, the goal is to estimate the most likely model
${\bf m}$. Invoking a Bayesian approach (and assuming a uniform prior), this 
can be turned to maximizing the likelihood function 
${\cal L} \equiv {\cal P}({\bf d}|{\bf m})$, the probability of
the data given the model, as a function of the model parameters.
Under the assumption that both the underlying velocities and the
observational errors are Gaussian random fields, the likelihood
function can be written as 
$ {\cal L}  =  [ (2\pi)^N \det(R)]^{-1/2} 
\exp\left( -{1\over 2}\sum_{i,j}^N {d_i R_{ij}^{-1} d_j}\right), $
where $\{d_i\}_{i=1}^{N}$ is the set of observed peculiar
velocities and $R$ is their correlation matrix. $R$ involves the theoretical 
correlation, calculated in linear theory for each assumed cosmological
model, and the estimated covariance of the errors.

The likelihood analysis is performed by choosing some parametric
functional form for the PS. Going over the parameter space and calculating 
the likelihoods for the different models, one finds the PS parameters for 
which the maximum likelihood is obtained. Note that this method, based on 
velocities, in fact measures $P(k){\Omega_m}^{1.2}$ and not the PS by itself. 
Confidence levels are estimated by approximating $-2{\rm ln}{\cal L}$ as a 
$\chi^2$ distribution with respect to the model parameters. 
We have extensively tested the reliability of the method on realistic mock 
catalogs, designed to mimic in detail the real data.

Our main analysis is done with a suite of generalized CDM models, normalized
by the COBE 4-year data. These include open models,
flat models with a cosmological constant and tilted models with and
without a tensor component, where the free parameters are the mass-density 
parameter $\Omega_m$, the Hubble parameter $h$ and the power index $n$.
We also use CDM-like models where the amplitude was allowed to vary.

A common problem in PS estimations is that the recovered PS is 
sensitive to the assumed observational errors, that contribute as well to
the correlation matrix $R$. To alleviate this problem, we extend the 
method so that also the magnitude of these errors is determined by the 
likelihood analysis. This is done by adding free parameters that 
govern global changes of the assumed errors, in addition to modeling 
the PS, and provides a consistency check of the magnitude of the 
errors. We find, for both catalogs, a good agreement with the original 
error estimates, which allows for a more reliable recovery of the PS.

\begin{figure}
\vspace{-4cm}
\plotone{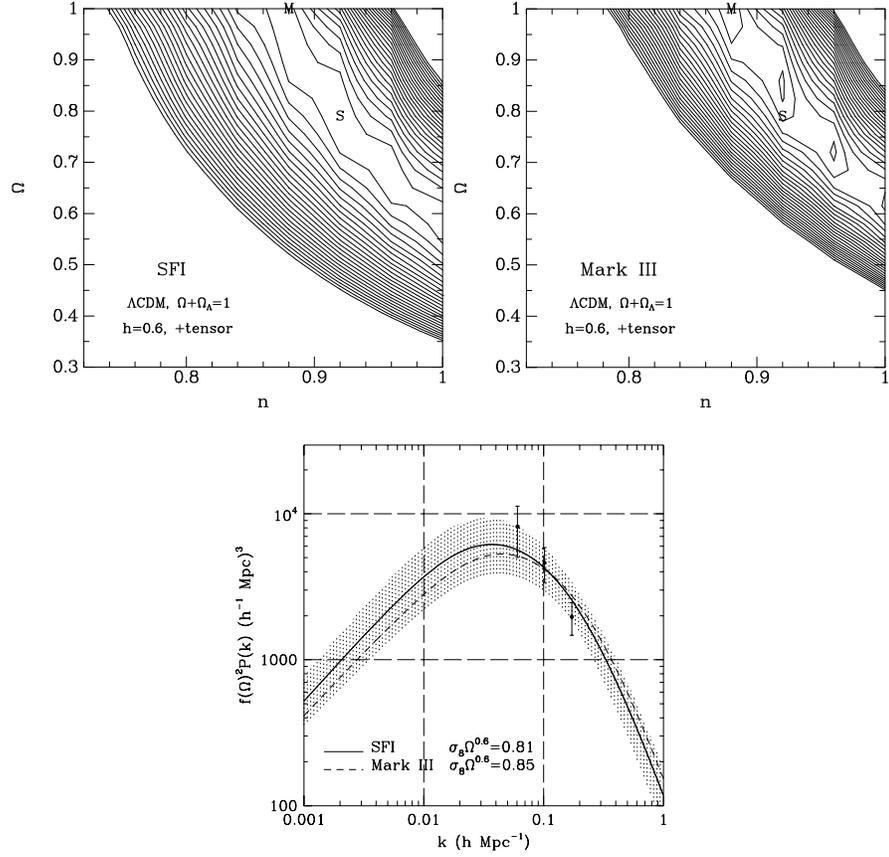}
\vspace{-3 cm}
\caption{Likelihood analysis results for the COBE-normalized flat
$\Lambda$CDM model with $h=0.6$. Shown are ${\rm ln}{\cal L}$ contours in the 
$\Omega_m-n$ plane for SFI (top left panel; cf.\ Freudling et al.\ 1999) 
and for Mark~III (top right; cf.\ Zaroubi et al.\ 1997). The best-fit 
parameters for SFI and Mark~III are marked on both by `S' and `M',
respectively. The lower panel shows the corresponding maximum-likelihood 
PS for SFI (solid line) and for Mark~III (dashed). The shaded region is 
the SFI $90\%$ confidence region. The three solid dots mark the PS calculated 
from Mark~III by Kolatt and Dekel (1997), together with their quoted  
$1\sigma$ error-bar. (in the plot $\Omega\equiv\Omega_m$) }
\label{fig:sfi_mark}
\end{figure}

\section{Results}
\label{sec:res}

Figure~\ref{fig:sfi_mark} shows, as a representative example, the results 
for the COBE-normalized flat $\Lambda$CDM family of models, with a tensor 
component in the initial fluctuations, when setting $h=0.6$ and varying 
$\Omega_m$ and $n$. 
Shown are ${\rm ln}{\cal L}$ contours for the SFI catalog and for Mark~III. As 
can be seen from the elongated contours, what is determined well is not 
a specific point but a high likelihood ridge, constraining a degenerate 
combination of the parameters roughly of the form $\Omega_m\, n^{3.7} = 
0.6 \pm 0.1$, in this case. The corresponding best-fit PS for the two 
catalogs is presented as well, with the shaded region illustrating the 
$90\%$ confidence region obtained from the SFI high-likelihood ridge.

These results are typical for all other PS models we tried. For
each catalog, the different models (including the ones with free 
normalization) yield similar best-fit PS, falling well within each others 
formal uncertainties and agreeing especially well on intermediate scales 
($k \sim 0.1 \, h\,{\rm Mpc^{-1}}$). The similarity of the PS obtained
from SFI with that of Mark~III, which is apparent in the figure, is 
also illustrative of the other models. This indicates that the
peculiar velocities of the two catalogs, with their respective error 
estimates, are consistent with arising from the same underlying mass 
density PS. This does not preclude possible differences 
that are not picked up by this statistic, but can be viewed as another
indication of the robustness of the results. Note also the agreement 
with an independent measure of the PS from the Mark~III catalog, using 
the smoothed density field recovered by POTENT (the three dots; Kolatt 
\& Dekel 1997).

More details on the method and results can be found in Zaroubi et al.\ 
(1997) and Freudling et al.\ (1999), regarding the applications to 
Mark~III and to SFI, respectively.
The robust result, for both catalogs and all models, is a relatively
high PS, with e.g.\ $P(k=0.1\, h\,{\rm Mpc^{-1}}){\Omega_m}^{1.2} = (4.5\pm2.0)\times 10^3 \, ( h^{-1} {\rm Mpc})^3$. 
An extrapolation to smaller scales using the different
CDM models gives $\sigma_8 {\Omega_m}^{0.6} = 0.8 \pm 0.2$. 
The high-likelihood ridge is a feature of all COBE-normalized CDM models,
corresponding to a general constraints on the combination of cosmological 
parameters of the sort $\Omega_m\, {h_{60}}^\mu\, n^\nu = 0.6 \pm 0.2$, where
$\mu = 1.3$ and $\nu = 3.7,\ 2.0$ for flat $\Lambda$CDM models with and 
without tensor fluctuations respectively. For open CDM, without tensor
fluctuations, the powers are $\mu = 0.9$ and $\nu = 1.4$.  
These error-bars are crude, reflecting the $90\%$ formal likelihood
uncertainty for each model, and the variance among different models and
catalogs. Note that our results are suggestive of somewhat higher values 
of $\sigma_8 {\Omega_m}^{0.6}$ and $\Omega_m$ than those implied by 
some other methods (such as from the clusters abundance and the
different $\beta$ measures).

\section{$\Omega_m - \Omega_\Lambda$ Constraints}
\label{sec:om_lam}

We have recently extended the analysis of COBE-normalized CDM models
to models with general values of $\Omega_m$ and $\Omega_\Lambda$ 
(Zehavi \& Dekel 1999). Although the $\Omega_\Lambda$ dependence comes 
in only indirectly through the COBE normalization, such results are 
particularly interesting as they can be combined with other constraints in
the $\Omega_m - \Omega_\Lambda$ plane.

\begin{figure}
\plotone{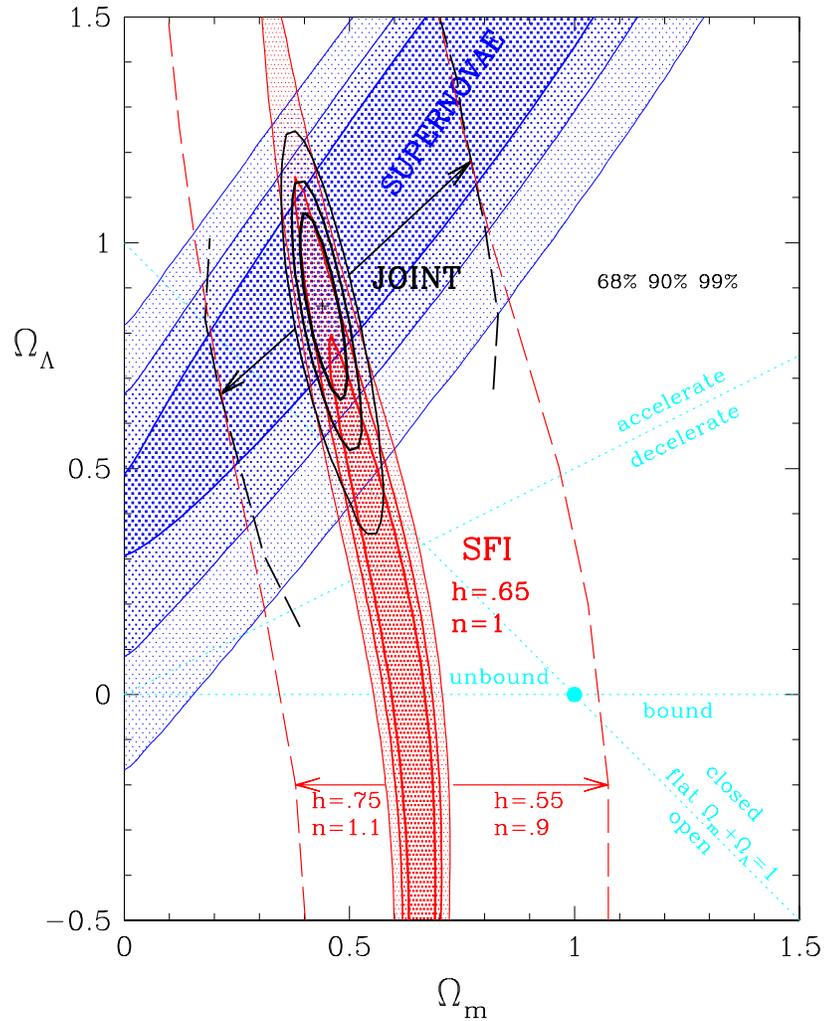}
\vspace{-3 cm}
\caption{Constraints in the $\Omega_m - \Omega_\Lambda$ plane arising from 
the high-z SN (the blue inclined contours; Perlmutter et al.\ 1999) and 
from a likelihood analysis of the SFI peculiar velocities (the red roughly 
vertical contours). The $68$, $90$ and $99\%$ confidence regions are shown 
for both. The peculiar velocity contours are for $n=1$, $h=0.65$, with the 
shifted red dashed lines showing the effect of changing the values of
these parameters. The corresponding joint confidence regions of the
velocities and the SN are shown as the black overlapping ellipses (Zehavi \& 
Dekel 1999). } 
\label{fig:om_lam}
\end{figure}

Figure~\ref{fig:om_lam} illustrates constraints in the 
$\Omega_m-\Omega_\Lambda$ plane, showing the confidence contours
obtained  by the Supernova Cosmology Project (Perlmutter et al.\ 1999;
consistent with the findings of the High-z Supernova Search Team, Riess 
et al.\ 1998) as well as the contours from our SFI likelihood analysis, 
performed for fixed values of $n=1$ and $h=0.65$. 
The velocity analysis constrains an elongated ridge in this plane of a 
nearly fixed $\Omega_m$ and varying $\Omega_\Lambda$. 
The analogous constraints from the Mark~III catalog are quite similar, 
except that the (fairly uncertain) upper bounds on $\Omega_\Lambda$ 
are slightly tighter. 
A change in the values of $n$ and $h$ essentially 
shifts the ridge toward a higher or lower $\Omega_m$, for smaller and 
larger values of these parameters, respectively. (This is another 
manifestation of the degeneracy between these parameters mentioned earlier). 
Their acceptable range is therefore needed to be determined a-priori from 
other constraints. For a reasonable range of these parameters, the effect 
is denoted in the plot by the shifted dashed lines.

The corresponding joint contours of the velocities and the SN, obtained 
by multiplying the likelihoods, are also shown on the plot, and here 
as well one should consider the area bounded by the dashed lines. 
Taking into consideration concurrently these two independent sets of 
constraints thus implies a considerable contribution from both 
$\Omega_m$ and $\Omega_\Lambda$. Specifically, models with small $\Omega_m$
and small $\Omega_\Lambda$, that are still allowed by the SN
constraints alone, are disfavored when the constraints from SNe and 
peculiar velocities are considered jointly; together, they make
a stronger case for an unbound accelerating universe with a positive 
cosmological constant. 

\section{Further Analysis}
\label{sec:joint}

\begin{figure}
\vspace{-3cm}
\plotone{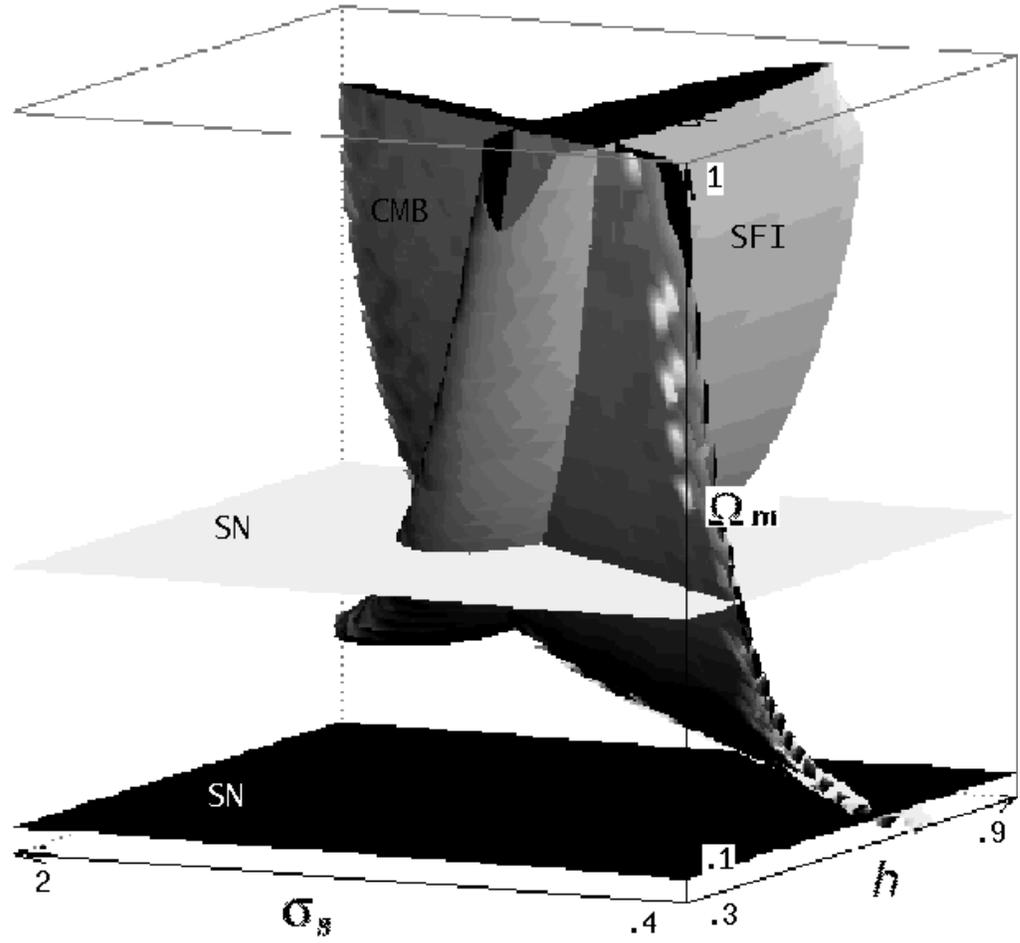}
\vspace{1cm}
\caption{The $95\%$ confidence regions obtained from the SFI velocities, 
the CMB anisotropies and the high-z SN, in the $\sigma_8 - h - \Omega_m$ 
space, for scale-invariant flat $\Lambda$CDM models (Bridle et al.\ 1999). }
\label{fig:joint}
\end{figure}

We are currently in the final stages of performing a joint analysis of peculiar
velocities together with constraints obtained from CMB anisotropies
and from the high-z SN (Bridle et al.\ 1999). The distinct types of data 
complement one another, each constraining different combinations of the 
cosmological parameters, and together can potentially set tight constraints. 
Figure~\ref{fig:joint} illustrates the constraints obtained for each of 
these data sets in the $\sigma_8 - h - \Omega_m$ space, for the 
scale-invariant {\it flat} $\Lambda$CDM family of models. 
The results shown here are for the SFI catalog.
Since the COBE constraint is now included in the CMB data,
the velocity analysis is performed free of the COBE normalization.
The high-z SN constraints appear here as bounds
on $\Omega_m$ (as we assume flat $\Lambda$CDM). 
The value of $h$ is left free to be determined by the likelihood analysis. 

The figure demonstrates that these three data sets provide roughly 
orthogonal constraints, with a significant overlap beyond the $2\sigma$ 
level, that allows a meaningful joint parameter estimation. 
The best-fit values obtained from the joint analysis of the three data sets 
are $\Omega_m=0.52$, $h=0.57$ and $\sigma_8=1.1$ (corresponding to 
$\sigma_8 {\Omega_m}^{0.6}=0.74$). The $95\%$ confidence regions on the
individual parameters, obtained by marginalizing over the other two
parameters, are $0.27 < \Omega_m < 0.54$, $0.54 < h < 0.85$ and
$1.02 < \sigma_8 < 1.67$. See also Lahav, these proceedings, for a
presentation of the results in the $\sigma_8 {\Omega_m}^{0.6} - \Omega_m h$
plane.

Additional work in progress includes an application of the likelihood 
method to other peculiar-velocity data, such as velocities of galaxy 
clusters (e.g., the SMAC sample; Hudson et al.\ 1999, Smith, these 
proceedings) or the local SN velocities, both probing 
larger scales with relatively high accuracy per object.  With 
regard to the cluster velocities, we are attempting to properly take 
into account in the analysis the fact that they are sampled at high 
density regions (Zehavi et al., in prep.).

Finally, another project underway is an attempt to obtain 
model-independent band-power estimates of the PS from peculiar velocities, 
using an iterative quadratic estimator scheme, 
which greatly improves the computational effort. 
(Such an approach is commonly applied to CMB measurements, e.g Bond, 
Jaffe, \& Knox 1998). This allows to relax the a-priori assumption of the 
PS form, and would illuminate the actual constraints on the different scales 
and their cross-correlations (Zehavi \& Knox 1999). 

\acknowledgments
We thank our collaborators in different aspects of the work presented here: 
S.\ Bridle, L.N.\ da Costa, A.\ Eldar, W.\ Freudling, R.\ Giovanelli, 
M.P.\ Haynes, Y.\ Hoffman, T.\ Kolatt, O.\ Lahav, J.J.\ Salzer, 
G.\ Wegner and S.\ Zaroubi. 
This research was supported by US-Israel Binational Science Foundation
grant 95-00330 and Israel Science Foundation grant 546/98 at HU, and by
the DOE and the NASA grant NAG 5-7092 at Fermilab.

\end{document}